\documentclass[]{tJOT2e}
\usepackage{hyperref}
\usepackage{natbib}

\usepackage{epstopdf}
\usepackage{color,ulem}

\begin{document}


 \markboth{E.~V.~Votyakov and S.~C.~Kassinos}{Journal of Turbulence}


\title{Core of the Magnetic Obstacle}

\author{
    E.~V.~Votyakov$^{\rm a}$$^{\ast}$\thanks{$^\ast$Corresponding author.
    Email: karaul@gmail.com \vspace{6pt}} and S.~C.~Kassinos$^{\rm a}$
    \\\vspace{6pt}
    $^{\rm a}$Computational Science Laboratory UCY-CompSci\\
    Department of Mechanical and Manufacturing Engineering,\\
    University of Cyprus, \\
    75 Kallipoleos, Nicosia 1678, Cyprus\\}

\maketitle

\begin{abstract}

Rich recirculation patterns have been recently discovered in the
electrically conducting flow subject to a local external magnetic
termed ``the magnetic obstacle" [Phys. Rev. Lett. 98 (2007),
144504]. This paper continues the study of magnetic obstacles and
sheds new light on the core of the magnetic obstacle that develops
between magnetic poles when the intensity of the external
 field is very large. A series of both 3D and 2D numerical
simulations have been carried out, through which it is shown that
the core of the magnetic obstacle is streamlined both by the
upstream flow and by the induced cross stream electric currents,
like a foreign insulated insertion placed inside the ordinary
hydrodynamic flow. The closed streamlines of the mass flow
resemble contour lines of electric potential, while closed
streamlines of the electric current resemble contour lines of
pressure. New recirculation patterns not reported before are found
in the series of 2D simulations. These are composed of many (even
number) vortices aligned along the spanwise line crossing the
magnetic gap. The intensities of these vortices are shown to
vanish toward to the center of the magnetic gap, confirming the
general conclusion of 3D simulations that the core of the magnetic
obstacle is frozen. The implications of these findings for the
case of turbulent flow are discussed briefly.

\bigskip


\bigskip
\bigskip

\end{abstract}

\section{Introduction}
\begin{figure*}[t]
\begin{center}
    \includegraphics[width=13.5cm, angle=0, clip]{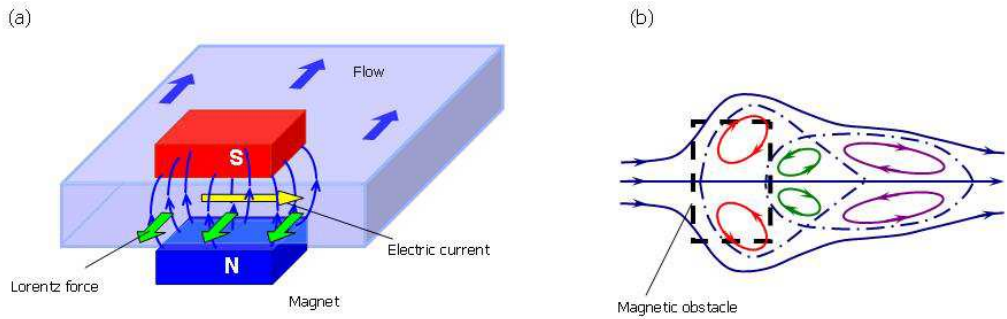}
\end{center}
    \caption{\label{Fig:IntroductoryMO}$a$ - scheme of the magnetic obstacle
    created by two permanent magnets which are located on the top and bottom of the channel where
    an electrically conducting liquid flows.  $b$ - structure of the wake of the magnetic
    obstacle consisting of  inner magnetic (first pair), connecting (second) and attached
    vortices (third pair). Dashed bold lines on $b$ mark borders of the
    magnets. }
\end{figure*}

A magnetic obstacle is a region in the flow of an electrically
conducting fluid, e.g. liquid metal, where an external
inhomogeneous magnetic field, ${\bf B}$, is applied as shown in
Fig.~\ref{Fig:IntroductoryMO}$a$. The region of the magnetic
obstacle manifests itself through the braking Lorentz force, ${\bf
F_L=j\times B}$, originating from the interaction of ${\bf B}$
with electrical currents ${\bf j}$. The electrical currents are
induced because of the electromotive force arising when the
conducting liquid moves through the region of magnetic field. The
net effect is that the core of the magnetic obstacle is
impenetrable to the flow, much like a foreign solid insertion.

Characteristics of the flow influenced by a magnetic obstacle are
of considerable fundamental and practical interest. On the
fundamental side, such a system possesses a rich variety of
dynamical states \cite{Votyakov:PRL:2007}. On the practical side,
spatially localized magnetic fields enjoy a variety of industrial
applications in metallurgy, e.g. \cite{Davidson:Review:1999},
including stirring of melts by a moving magnetic obstacle (called
electromagnetic stirring), removing undesired turbulent
fluctuations during steel casting using steady magnetic obstacles
(called electromagnetic brake) and non-contact flow measurement
using a magnetic obstacle (called Lorentz force velocimetry, e.g.
\cite{Thess:Votyakov:Kolesnikov:2006}).

In this paper, the magnetic Reynolds number $R_m=\mu^*\sigma u_0
H$ is taken to be much less than one where $\mu^*$ is the magnetic
permeability, $\sigma$ is fluid electric conductivity, and $u_0$
and $H$ are the characteristic scales for velocity and length.
Therefore the induced magnetic field is expected to be much less
that the imposed external magnetic field
\cite{Shercliff:book:1962}, \cite{Roberts:1967},
\cite{Moreau:book:1990}, \cite{Davidson:book:2001}. Under this
constraint, the external magnetic field has the following twofold
effect on a turbulent magnetohydrodynamic (MHD) flow. Firstly, the
turbulent velocity pulsations are suppressed in the direction
parallel to the direction of the external field, that is the
turbulence tends to be more and more two-dimensional when the
external field becomes stronger and stronger \cite{Moffatt:1967},
\cite{Sommeria:Moreau:MHDturblulence:1982}, \cite{Davidson:1997}.
This is true for the system subject to a homogeneous magnetic
field where the mean velocity is constant over the flow except for
boundary layers. Secondly, when the external field is local in
space, as it must be in the case of the magnetic obstacle, then
the decelerating Lorentz force is higher in the center of the
obstacle compared to its periphery. This creates a shear gradient
in the mean flow velocity which generates an additional vorticity
which then diffuses downstream and contributes to the turbulence.
Therefore it is important to understand for practical applications
whether the useful turbulence-damping effect of a magnetic brake
is not obliterated by excessive vorticity generation in the wake
of the magnetic obstacle.

In order to deal with turbulent phenomena one needs to know the
averaged parameters of the flow. However to properly define, for
instance, the mean velocity is not a trivial task in the case of a
local magnetic field because various recirculation patterns are
possible both inside and in the vicinity of the magnetic obstacle
as shown below. So the first obvious step is to study a laminar
flow around the magnetic obstacle before attacking turbulence
subject to the local external magnetic field. On the other hand,
as shown in this paper and elsewhere \cite{Votyakov:PoF:2009},
there is a similarity concept between a hydrodynamic flow around a
solid cylinder and a MHD flow around a strong magnetic obstacle.
This gives hope that numerous results for turbulent flows
initiated by an obstacle can be projected onto turbulent MHD flows
influenced by the local magnetic field. For instance, the
vorticity generation by shear layer of a solid cylinder can be
roughly perceived as similar to those in the shear layer alongside
the magnetic obstacle. Thus, this paper is aimed to attract
attention of researchers working on ordinary hydrodynamic
turbulence to problems appearing in a MHD flow subject to a
heterogeneous external magnetic field.

Studies of the effects of the magnetic obstacle on a liquid metal
flow had been initiated in 1970s in the former Soviet Union by
Gelfgat \textit{et al.} \cite{Gelfgat:Peterson:Sherbinin:1978},
\cite{Gelfgat:Olshanskii:1978}, and have been recently revived in
the West by Cuevas \textit{et al.}
\cite{Cuevas:Smolentsev:Abdou:Pamir:2005},
\cite{Cuevas:Smolentsev:Abdou:2006},
\cite{Cuevas:Smolentsev:Abdou:PRE:2006}. Among the above citations
there were 2D numerical works related to creeping MHD flow, where
a possible recirculation induced by the local magnetic was shown,
see for example \cite{Gelfgat:Peterson:Sherbinin:1978},
\cite{Cuevas:Smolentsev:Abdou:PRE:2006}, but where the physical
explanation of the recirculation, as well as the generic scenario
for the MHD flow around the magnetic obstacle, were obscured.

New results about the wake of a magnetic obstacle have been
reported by \cite{Votyakov:PRL:2007}, \cite{Votyakov:JFM:2007} and
the generic scenario has been elaborated in
\cite{Votyakov:PoF:2009}. It has been found that a liquid metal
flow subject to a local magnetic field shows different
recirculation patterns: (1) no vortices, when the viscous forces
prevail at small Lorentz force, (2) one pair of \textit{inner
magnetic} vortices between the magnetic poles, when Lorentz force
is high and inertia small, and (3) three pairs, namely, magnetic
as above, \textit{connecting} and \textit{attached} vortices, when
Lorentz and inertial forces are high. The latter six-vortex
ensemble is shown in Fig.~\ref{Fig:IntroductoryMO}$b$.

The goal of the current paper is to highlight effects taking place
in the laminar flow around the magnetic obstacle when the
interaction parameter $N$, which is a ratio of Lorentz force to
the inertial force, increases. When $N$ is very large, both mass
transfer and electric field vanish in the region between magnetic
poles. This region, hereinafter called the core of the magnetic
obstacle, appears as if frozen by the external magnetic field, so
that the upstream flow and crosswise electric currents can not
penetrate inside it. Thus, the core of the magnetic obstacle is
similar to a solid insulated obstacle inside an ordinary
hydrodynamical flow with crosswise electric currents and
\textit{without} an external magnetic field. (This concerns
hydrodynamics because there is no magnetic field, and the
crosswise electric currents go around the insulated insertion
without changing the mass flow.) Magnetic vortices are located
aside the core and compensate shear stresses like a ball-bearing
between the impenetrable region and upstream flow.

It is worthwhile to notice that the core of a magnetic obstacle is
not similar to a stagnant region appearing in the MHD flow subject
to a fringing transverse magnetic field. The fringing field is
nonuniform in the streamwise direction and uniform in the spanwise
direction, while in the case of the  magnetic obstacle, the
external local magnetic field is nonuniform in all the directions.
Effects of the fringing field on a duct liquid metal flow have
been intensively studied before, see for example
\cite{Molokov:Reed:Fusion:2003}, \cite{Alboussiere:2004},
\cite{Kumamaru:etal:2004}, \cite{Kumamaru:etal:2007},
\cite{ni_current_2007}, while those of the magnetic obstacle is a
new research field \cite{Votyakov:PoF:2009}. It is trivial that in
both cases the breaking Lorentz force is responsible for the
phenomenon that the liquid metal flow vanishes in the space of the
strong magnetic field. However, in the case of the fringing field
the side flow jets are caused by a geometrical heterogeneity
imposed by the sidewalls of the duct, and the stagnant region of
vanishing flow tends to spread completely between the sidewalls.
In the case of the magnetic obstacle, maxima of streamwise
velocity appear in an originally free flow around the region where
the magnetic field is of highest intensity, and the core roughly
corresponds to the region where the magnetic field is imposed. The
uniformity of the fringing field in the spanwise directions makes
impossible recirculation in the M-shaped velocity profile, while
magnetic vortices alongside a magnetic obstacle easily appear at
moderate $N$ \cite{Votyakov:JFM:2007}.

The structure of the present paper is as follows. First, we
present technical details of the simulations: model, equations and
3D numerical solver. Then, we report results for the core of the
magnetic obstacle obtained in a series of 3D simulations. As an
extension of the presentation given in TSFP-6
\cite{Votyakov:TSFP6:2009}, we discuss also a 2D MHD flow and show
the differences between 2D and 3D cases. A summary of the main
conclusions ends the paper.

\section{3D Model: equations and numerical method}
The equations governing the motion of  an electrically conductive
and incompressible fluid are derived from the Navier-Stokes
equations coupled with the Maxwell equations for a moving medium
and Ohm's law. By assuming that the induced magnetic field is
infinitely small in comparison to the external magnetic field, the
equations in dimensionless form are:
\begin{eqnarray}
    \label{eq:NSE:momentum}
    \frac{{\partial\textbf{u}}}{{\partial t}} + (\textbf{u} \cdot \nabla ) \textbf{u}
    &=& - \nabla p + \frac{1}{Re}\triangle \textbf{u} + N (\textbf{j}\times\textbf{B}),
    \\ \label{eq:NSE:Ohm} \textbf{j} &=&  -\nabla\phi + \textbf{u} \times \textbf{B},
    \\ \label{eq:NSE:Poisson}  \nabla \cdot \textbf{j} &=& 0,
    \\ \label{eq:NSE:continuity} \nabla \cdot \textbf{u} &=& 0,
\end{eqnarray}where $\textbf{u}$ is  velocity field, $\bf{B}$ is an
external magnetic field, $\textbf{j}$ is electric current density,
$p$ is pressure, $\phi$ is electric potential.  The Reynolds
number, $Re=u_0H/\nu$, expresses a ratio between the inertia force
and the viscous force and the interaction parameter,
$N=B_0^2H\sigma/(u_0\rho)$, expresses the ratio between the
Lorentz force and the inertia force. $Re$ and $N$ are linked with
each other by means of the Hartmann number, $Re\,N=Ha^2$,
$Ha=HB_0(\sigma/\rho\nu)^{1/2}$, which determines the thickness
$\delta$ of Hartmann boundary layers, $\delta/H \sim Ha^{-1}$,
formed near the walls perpendicular to the direction of the
magnetic field in the flow under constant magnetic field. Here,
$H$ is the characteristic length (size), $u_0$ is the
characteristic flow velocity, $B_0$ the characteristic magnitude
of the magnetic field intensity, $\nu$ is the kinematic viscosity
of the fluid, $\sigma$ is the electric conductivity of the fluid,
and $\rho$ is its density.

At given $Re$, $N$ and ${\bf B}(x,y,z)$, the system of the partial
differential equations shown above is solved in a 3D computational
domain to obtain the unknown  ${\bf u}(x,y,z)$, $p(x,y,z)$ and
$\phi(x,y,z)$. The computational domain has periodical boundary
conditions in the spanwise direction, and no-slip and insulating
top and bottom walls (transverse direction). The electric
potential at the inlet and outlet planes is taken equal to zero.
For the velocity, the outlet boundary is force free, and a laminar
parabolic velocity profile is imposed at the inlet boundary. We
are interested in a stationary laminar solution, hence, the
initial conditions play no role.

The origin of the right-handed coordinate system, $x=y=z=0$, is
taken in the center of the magnetic gap. The size of the
computational domain is: $-L_{x}\le x \le L_{x}$, $-L_{y}\le y \le
L_{y}$, $-H \le z \le H$, where $L_{x}=25$, $L_{y}=25$, $H=1$ and
$x, y, z$ are respectively the streamwise, crosswise, and
transverse directions.

The characteristic dimensions for the Reynolds number $Re$, and
the interaction parameter $N$ are the half-height of the duct $H$,
the mean flow rate $u_0$, and the magnetic field intensity $B_0$
taken at the center of the magnetic gap, $x\!=\!y\!=\!z\!=\!0$.
The range of the studied parameters is: $Re=0.1, 1, 10, 100$ and
$0 \leq N \leq 1000$.

The external magnetic field is modelled as a field from two
permanent magnets occupying a space $\Omega=\{|x|\leq M_x, |y|\leq
M_y, |z|\geq h\}$, where $M_x=1.5$ ($M_y=2$) is the streamwise
(spanwise) width of the magnet, and $2\times h$ is the distance
between magnetic poles, $h=1.5$. The magnets are supposed to be
composed of magnetic dipoles oriented along the $z$-direction,
therefore the total magnetic field  ${\mathbf
B(x,y,z)}=\int_{\Omega}{\bf B_d(r,r')} d{\bf r'}$, where ${\bf
B_d(r,r')}=\nabla\left[\partial_{z}(1/|{\bf r}-{\bf r'}|)\right]$
is a field, at the point $\mathbf{r}=(x,y,z)$ created by the
single magnetic dipole located in the point
$\mathbf{r'}=(x',y',z')$. The integration can be performed
analytically, see \cite{Votyakov:JFM:2007}, and after cumbersome
algebraic calculations one obtains:
\begin{eqnarray*} \label{eq:MF:final}
    B_{x}(\mathbf{r})&=& \frac{1}{B_{0}} \sum_{k=\pm 1} \sum_{j=\pm 1}\sum_{i=\pm 1}
    (ijk)\,\mbox{arctanh}\left[\frac{\delta_j}{\delta_{ijk}}\right], \label{eq:MF:Bx}\\
    B_{y}(\mathbf{r})&=& \frac{1}{B_{0}} \sum_{k=\pm 1} \sum_{j=\pm 1}\sum_{i=\pm 1}
    (ijk)\,\mbox{arctanh}\left[\frac{\delta_i}{\delta_{ijk}}\right], \label{eq:MF:By}\\
    B_{z}(\mathbf{r})&=& -\frac{1}{B_{0}} \sum_{k=\pm 1} \sum_{j=\pm 1}\sum_{i=\pm
    1}(ijk)\mbox{arctan}\left[\frac{\delta_i \delta_j}{\delta_k\delta_{ijk}}\right], \label{eq:MF:Bz}
\end{eqnarray*} where $\delta_i=(x-iM_x)$, $\delta_j=(y-jM_y)$,
$\delta_k=(z-kh)$, and
$\delta_{ijk}=[(x-iM_x)^2+(y-jM_y)^2+(z-kh)^2]^{1/2}$. The
normalization factor $B_{0}$ is selected in such a way to have the
intensity of the $z$-component equal one, $B_z(0,0,0)=1$, in the
center of the magnetic gap. Three-fold summation with the
sign-alternating factor $(ijk)$ reflects the fact that these
equations are obtained by integrating over the 3D box $\Omega$.
Different cuts of the intensity ${\bf B(r)}$ are plotted in Fig.~3
and Fig.~4($b$) in the paper of \cite{Votyakov:JFM:2007}.

The 3D numerical solver has been explained in detail earlier, see
\cite{Votyakov:Zienicke:FDMP:2006}. It was developed from a free
hydrodynamic solver originally created in the research group of
Prof.~M.~Griebel (\cite{Griebel:book:1995}). The solver employs
the Chorin-type projection algorithm and finite differences on an
inhomogeneous staggered regular grid. Time integration is done by
the explicit Adams-Bashforth method that has second order
accuracy. Convective and diffusive terms are implemented by means
of the VONOS (variable-order non-oscillatory scheme) method. The
3D Poisson equations are solved for pressure and electric
potential at each time step by using the bi-conjugate gradient
stabilized method (BiCGStab).

The numerical grid was regular and inhomogeneous, $N_x\times N_y
\times N_z=64^3$. The minimal horizontal step size in the region
of the magnetic gap was $\Delta x \simeq \Delta y \simeq 0.3$,
which means that a few dozens points were used for resolving the
inner vortices in the core of the magnetic obstacle. The minimal
vertical step size near the top and bottom (Hartmann) walls was
$\Delta z=0.005$. This corresponds to using three to five
($=(1/Ha)/\Delta z)$ points to resolve Hartmann layer at
$Ha=40-70$. To ascertain that the numerical resolution was
adequate, a few runs were performed with double the resolution and
no differences have been found.

\section{3D results\cite{Votyakov:PoF:2009}}

The goal of the simulations is to focus on the flow around a
magnetic obstacle at large interaction parameter $N$. In order to
achieve large $N=Ha^2/Re$, the simulations were started at a small
interaction parameter and $Ha$ was smoothly increased, while
keeping $Re$ constant. Several values of the Reynolds number were
studied, $Re=0.1, 1, 10, 100$, and no principal differences were
found at the same $N$. These low values of $Re$ imply low inertial
forces, therefore, only two-vortex patterns were produced, without
connecting and attached vortices.

In this Section, results are shown for the mid central plane,
where all vortex peculiarities can be distinctively visualized.
Nevertheless, it is necessary to note that the flow in the mid
plane is not two-dimensional. There is a secondary flow from and
into the mid plane towards and from the top and bottom walls. This
secondary flow is caused by the process of creation and
destruction of the Hartmann layers. 3D pictures of the vortices
have been drawn before \cite{Votyakov:JFM:2007} and will not be
considered here.

\begin{figure*}[]
\begin{center}
    \includegraphics[width=13.5cm, angle=0, clip]{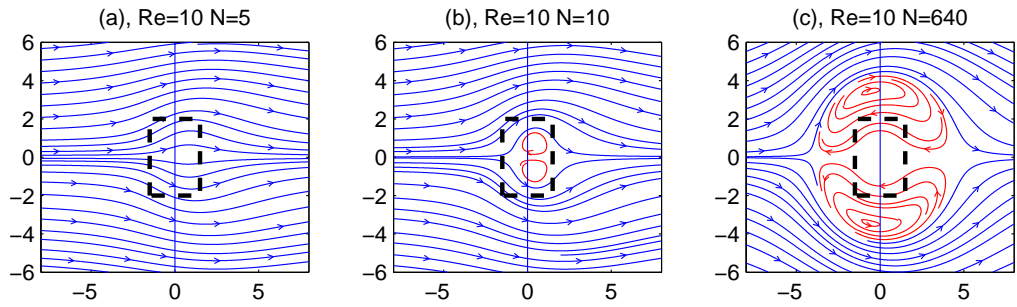}
\end{center}
    \caption{\label{Fig:StreamSlices}Streamlines in the central plane, $Re=10$,
    $N=$5($a$), 10($b$), 640($c$). Dashed bold lines mark borders of the
    magnets. As $N$ gives rise, magnetic vortices move away each other by
    forming in between a core of the magnetic obstacle.}
\end{figure*}

The natural way to visualize the core of the magnetic obstacle is
to plot streamlines of the flow in the central horizontal plane as
shown in Fig.~\ref{Fig:StreamSlices} at different interaction
parameters $N$.  Because $N$ is the ratio of the Lorentz force to
the inertial force, the larger $N$ is, the stronger the retarding
effect of the Lorenz force becomes. So, one observes no vortices
at $N=5$, Fig.~\ref{Fig:StreamSlices}$a$; weak circular magnetic
vortices first appear at slightly below $N=10$, as shown in
Fig.~\ref{Fig:StreamSlices}$b$; and finally these vortices become
well developed and strongly deformed at very large $N=640$,
Fig.~\ref{Fig:StreamSlices}$c$. In the latter case, the vortex
streamlines envelop the bold dashed rectangle. This rectangle
denotes the borders of the external magnet; inside the rectangle
at large $N$ one can see an island  --  the core of the magnetic
obstacle. The observed deformation of the vortices and their drift
from the center of the magnetic gap are due to the tendency of the
flow to reduce the friction caused by retarding Lorentz force. The
vortices are cambered and located in the shear layer alongside the
magnetic gap in such a way that their rotation looks like the
rotation of a ball-bearing inside the wheel.

\begin{figure*}[]
\begin{center}
    \includegraphics[width=13.5cm, angle=0, clip=on]{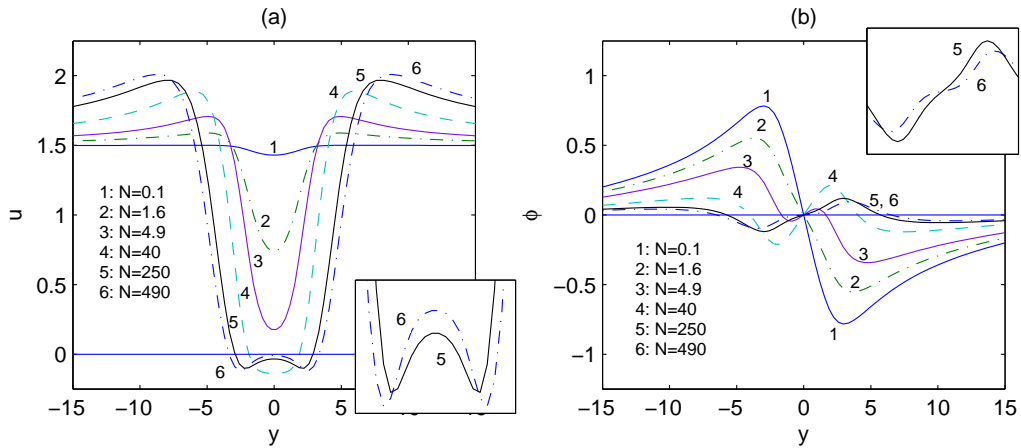}
\end{center}
    \caption{\label{Fig:u_phi_dist}Streamwise velocity ($a$) and electric
    potential ($b$) along crosswise cuts of middle horizontal plane
    $x=z=0$. $Re=10$, $N=$0.1(solid 1), 1.6(dot-dashed 2), 4.9(solid 3),
    40(dashed 4), 250(solid 5), and 490(dot-dashed 6). Insertion shows
    magnified plots for curves 5 and 6.}
\end{figure*}

\begin{figure*}[]
\begin{center}
    \includegraphics[width=13.5cm, angle=0, clip=yes]{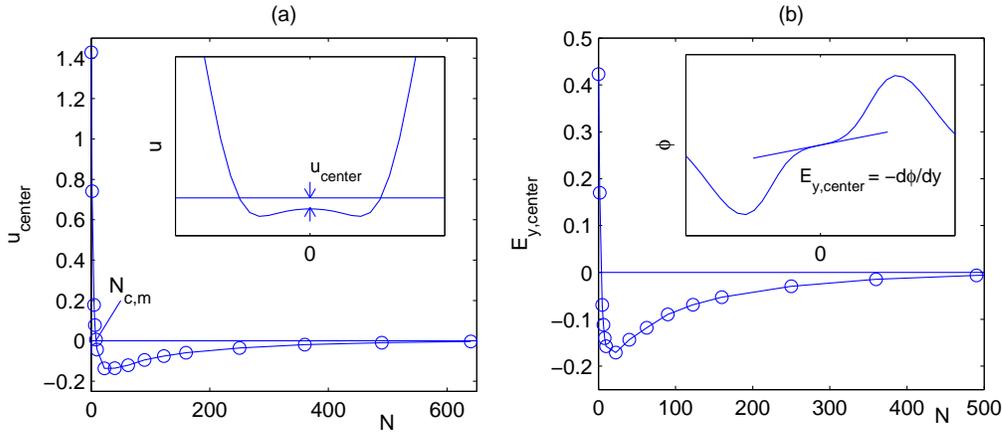}
\end{center}
    \caption{\label{Fig:u_phi_summ}Central streamwise velocity $u_{center}$ ($a$)
    and central spanwise electric field $E_{y,center}$ ($b$) as a function
    of the interaction parameter $N$. $N_{c,m}$ is a critical
    value where the streamwise velocity is equal to zero.
    Insertion shows the definition of $u_{center}$ and $E_{y,center}$.}
\end{figure*}

The quantitative analysis of the core was performed by crosswise
cuts through the center of the magnetic gap at different arising
magnetic interaction parameters $N$. These cuts are shown in
Fig.~\ref{Fig:u_phi_dist}a for the streamwise velocity $u_x(y)$
and in Fig.~\ref{Fig:u_phi_dist}b for the electric potential
$\phi(y)$. First, we discuss how the streamwise velocity changes
as $N$ increases.

As shown in the Fig.~\ref{Fig:u_phi_dist}$a$ (curve 1), for small
$N=0.1$, the velocity profile is only slightly disturbed with
respect to a uniform distribution. As $N$ increases, the curves
$u_x(y)$ pull further down in the central part $u_{center} \equiv
u_x(0)$, see for example curves 2 and 3. At $N$ higher than a
critical value $N_{c,m}$, i.e. for curve 4, the central velocities
$u_{center}$ are negative. This means that there appears a reverse
flow causing magnetic vortices in the magnetic gap. When $N$ rises
even more (see curves 5 and 6) the magnetic vortices become
stronger and simultaneously shift away from the center to the side
along the $y$ direction, see insertion in
Fig.~\ref{Fig:u_phi_dist}$a$ for curves 5 and 6.

Fig.~\ref{Fig:u_phi_dist}($b$) shows how the electric potential
$\phi(y)$ varies along the central crosswise cut through the
magnetic gap. The slope in the central point is the crosswise
electric field, $E_{y,center}=-d\phi/dy|_{y=0}$. One can see that
$E_{y,center}$ changes its sign:  it is positive at small $N$ and
negative at high $N$. To explain why it is so, one can use the
following way of thinking. Any free flow tends to pass over an
obstacle in such a way so as to perform the lowest possible
mechanical work, i.e. flow streamlines are the lines of least
resistance to the transfer of mass. The resistance of the flow
subject to an external magnetic field is caused by the retarding
Lorentz force $F_x\approx j_y B_z$, so the flow tends to produce a
crosswise electric current, $j_y$, as low as possible while
preserving the divergence-free condition $\nabla\cdot{\bf j}=0$.
To satisfy the latter requirement, an electric field ${\bf E}$
must appear, which is directed in such a way, so as to compensate
the currents produced by the electromotive force ${\bf
u}\times{\bf B}$. Next, we analyze the crosswise electric current
$j_y = E_y + (u_z B_x - u_x B_z)$. Due to symmetry in the center
of the magnetic gap $B_y=B_x=u_y=u_z=j_y=j_z=0$,  so $j_y = E_y -
u_x B_z$. This means that  $E_y$ tends to have the same sign as
$u_x$ in order to make $j_y$ smaller. At small $N$, the streamwise
velocity $u_x$ is large and positive, so the electric field $E_y$
is positive too. When the magnetic vortices appear, there is a
reverse flow in the center. Therefore, the central velocity is
negative now, and the central electric field $E_{y,center}$ is
also negative.

The change of the electric field in the magnetic gap can be
explained in terms of the Poisson equation and the concurrence
between external and internal vorticity, see
\cite{Votyakov:JFM:2007}. Those arguments are also valid here,
however, in contrast to the previous study, we have no side walls
now, so the external vorticity in the present case plays only a
minor role. As a result, the reversal of the electric field
appears at a small $N$ (approximately equal to five), which is
close to $\kappa=0.4$ given in \cite{Votyakov:JFM:2007}.

\begin{figure*}
\begin{center}
    \includegraphics[width=13.5cm, angle=0, clip=yes]{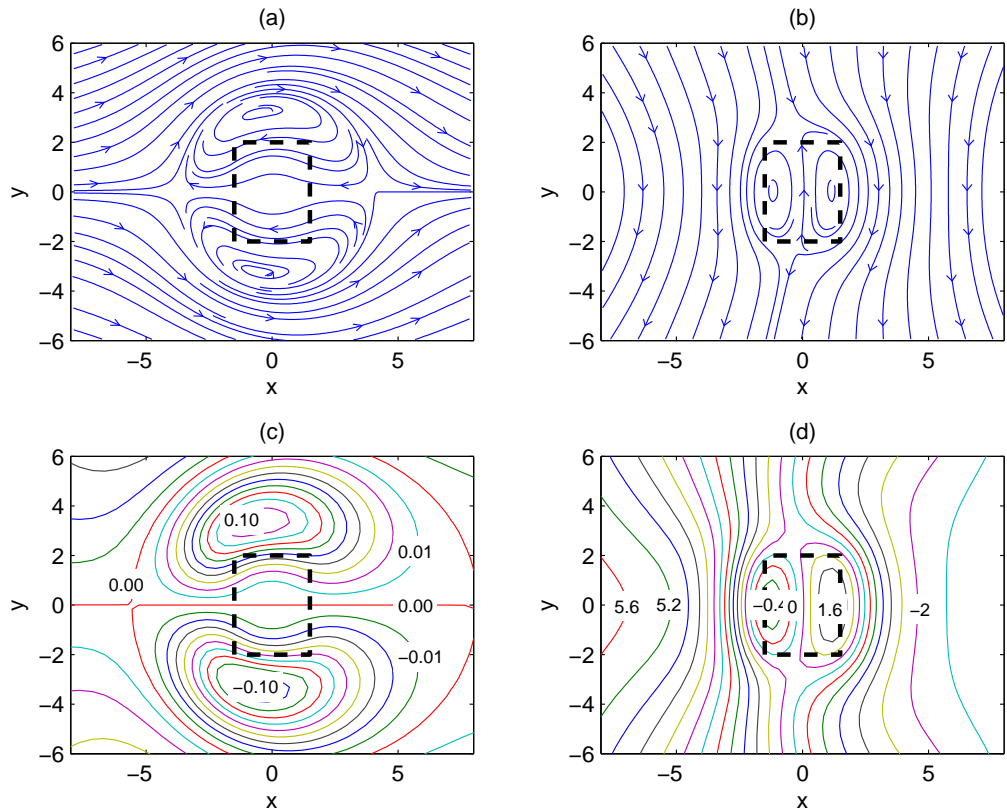}
\end{center}
    \caption{\label{Fig:phi_p_u_jRe0010Ha070}
    Middle horizontal plane, $z=0$:
    streamlines of the mass ($u_x, u_y$) ($a$)  and electric charge
    ($j_x, y_y$) ($b$) flow. Contour lines for the electric potential $\phi(x,y)$ ($c$)
    and pressure $p(x,y)$ ($d$) resemble the streamlines given above.
    $Re=10$,    $N=490$. Contours of the electric potential are given with step 0.01,
    and contours of the pressure are given with the step 0.4.
    Dashed bold rectangle shows borders of the external magnet.}
\end{figure*}

The overall data about $u_{center}$ and $E_{y,center}$ in the
whole range of $N$ studied are shown in Fig.~\ref{Fig:u_phi_summ}.
One can see that both characteristics start from positive values,
then, they cross the zeroth level, reach a minimum, go up again,
and finally vanish in the limit of high $N$. With respect to the
streamwise velocity, this means that, at hight $N$, there is no
mass flow in the center of the magnetic gap; the other velocity
components are equal to zero due to symmetry. With respect to the
crosswise electric field, this means that there are no electric
currents. This occurs because there is no mass flow, therefore,
the electromotive force vanishes, $E_y$ goes to zero, and the
other electric field components are equal to zero due to symmetry.
Thus, one can say that the center of the magnetic gap is frozen by
the strong external magnetic field, so that both mass flow and
electric currents tend to bypass the center. In other words, this
means that a strong magnetic obstacle has a core, and such a core
is like a solid insulated body, being impenetrable for the
external mass and electric charge flow.

When the inertia and viscous forces are negligible compared to the
Lorentz force and pressure gradients, then mass flow streamlines
must be governed by the electric potential distribution, while the
trajectories of the induced electric current must be governed by
pressure distribution. This is derived straightforwardly from
equations (\ref{eq:NSE:momentum} -- \ref{eq:NSE:Ohm}). Because
inertia and viscosity are vanishing, equations
(\ref{eq:NSE:momentum} -- \ref{eq:NSE:Ohm}), in the core and
nearest periphery of the magnetic obstacle, become:
\begin{eqnarray}
    \nabla p = {\bf j \times B}, \quad \nabla \phi = -{\bf j}+{\bf
    u\times B} \approx {\bf u\times B}\,. \label{eq:Kulikovskii}
\end{eqnarray}
In the latter formula, ${\bf j} \ll \nabla\phi \mbox{ and } {\bf
u\times B}$ is the dominating term. In the core of the obstacle,
${\bf B}=(0,0,B_z)\approx (0,0,1)$, hence, the pressure (electric
potential) is a streamline function for the electric current
(velocity), see Fig.~\ref{Fig:phi_p_u_jRe0010Ha070}. These
relationships for the flow under the strong external magnetic
field had been discussed earlier by Kulikovskii in 1968
\cite{Kulikovskii:1968}.

Kulikovskii's theory is linear, therefore, it must work well in
the stagnant core of the magnetic obstacle. The traditional
approach in the context of this theory is to introduce  the
so-called characteristic surfaces, and then, to impose Hartmann
layers as boundary conditions for further integration along the
characteristic surfaces. Such an approach has been used before for
slowly varying fringing magnetic fields \cite{Alboussiere:2004},
where  Hartmann layers and the inertialess assumption are
reasonable. However, it is an open issue whether the concept of
the characteristic surfaces is valid for the case of the magnetic
obstacle. For perfectly electrically conductive liquids this
concept forces mass and electric streamlines to flow along the
surfaces of constant $\mathbf{B}$. The latter is the conjecture of
(\ref{eq:Kulikovskii}) and is observed actually in
Fig.~\ref{Fig:phi_p_u_jRe0010Ha070}$a,b$ far from the core of the
obstacle. Nevertheless, the concept of  the characteristic
surfaces does not allow for any recirculation in shear layers,
what is the most remarkable effect here.

A magnetic field in a rotational flow requires more sophisticated
boundary conditions than just the Hartmann layer. There is known a
solution for the Ekman-Hartman layers \cite{Debnath:1973}, where
both constant rotation and constant magnetic field are taken
jointly into account. This probably does not fit the present case
either, because the vorticity is not constant along the transverse
direction, and the shape of vortices is not circular. Moreover,
inclusion of the non constant vorticity destroys the linearity of
Kulikovskii's theory. Therefore, Kulikovskii's theory could not be
used as it stands to predict recirculation \emph{a priori}.
Indeed, this explains why the theory has not been applied to
magnetic vortices, even though it has been known for a while.
Nevertheless, Kulikovskii's theory is useful and must be mentioned
because it explains \emph{a posteriori} the shape of vortices and
their matching to electric potential lines.

\section{2D creeping flow around magnetic obstacle}

It is interesting to compare our results with those reported
earlier by Cuevas \textit{et.al}
\cite{Cuevas:Smolentsev:Abdou:PRE:2006} for a creeping 2D MHD
flow. In particular, these authors observed not only two magnetic
vortices, but four vortices as well as, aligned in the spanwise
direction at high Hartmann numbers, see Fig.~9 of
\cite{Cuevas:Smolentsev:Abdou:PRE:2006}. The four-vortex aligned
structure has been missed in our 3D simulations, therefore to get
insight about it, we have performed our own 2D simulations with
the same parameters for the magnetic field $M_x=0.5$, $M_y=0.5$,
$h=1$ (our formula for the field (\ref{eq:MF:Bz}) coincides with
Eq.~(31) of \cite{Cuevas:Smolentsev:Abdou:PRE:2006}) and
$Re=0.05$. To gather more information, we explored the larger $Ha$
range, $0\le Ha \le 150$. A 2D finite element method\footnote{In
the 2D simulations, the numerical mesh varied from $64^2$ to
$256^2$.  The quality of the mesh was checked by repeating the
runs with doubled resolution.} was numerically employed to solve
the vorticity-stream function formulation of steady-state
Navier-Stokes equation with the Lorentz force. Electric currents
were calculated from the magnetic induction equation. Generally,
the obtained 2D results coincided with those from
\cite{Cuevas:Smolentsev:Abdou:PRE:2006} and also extended them as
reported below.

\begin{figure*}[]
\begin{center}
    \includegraphics[width=12cm, angle=0, clip=on]{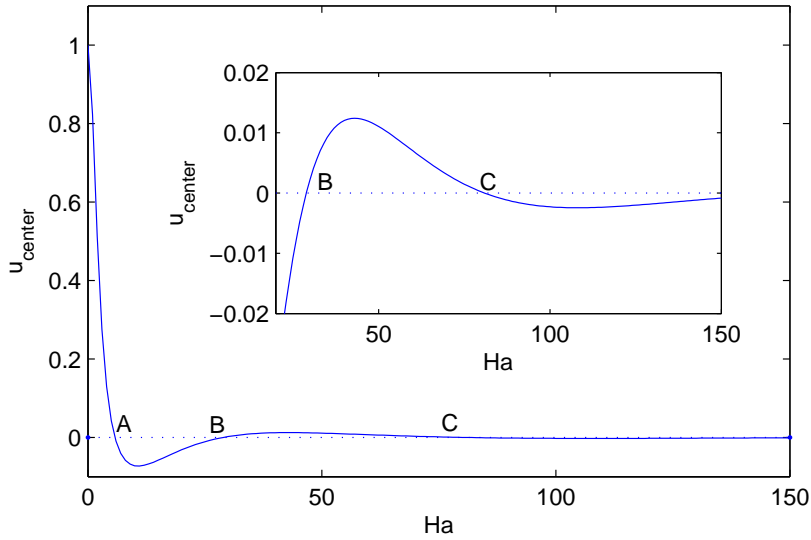}
\end{center}
    \caption{\label{Fig:u_summ_2DRe005}
    Central streamwise velocity $u_{center}$ as a function of $Ha$ for 2D MHD
    flow, $Re=0.05$, $M_x=0.5$, $M_y=0.5$, $h=1$.
    Insertion is the zoom for $20\le Ha\le 150$.}
\end{figure*}

The main curve of our 2D results is presented in
Fig.~\ref{Fig:u_summ_2DRe005}.  It shows the dependence of the
velocity in the center of the magnetic obstacle as a function of
$Ha$ number. (This is similar to Fig.~\ref{Fig:u_phi_summ}($a$)
with the difference that the $x$-axis corresponds to $Ha$ instead
of $N$.) One can see that $u_{center}$ vanishes as $Ha$ increases,
i.e. mass flow disappears in the core of the obstacle as the
intensity of the magnetic field gives rise. Therefore, the
principal conclusion derived from 3D results remains valid: there
develops a frozen core of the magnetic obstacle when  $Ha$ is
large. However, the details of the to approach zero-level values
for $u_{center}$ are different in 2D case relative to the 3D case.
To stress this fact,  we zoomed the curve in
Fig.~\ref{Fig:u_summ_2DRe005} and put it as an insertion. One can
see that in the 2D case $u_{center}$ vanishes with oscillations,
while in the 3D case, see Fig.~\ref{Fig:u_phi_summ}($a$), it was
always negative as it approaching the zero level.

\begin{figure*}[]
\begin{center}
    \includegraphics[width=12.5cm, angle=0, clip=on]{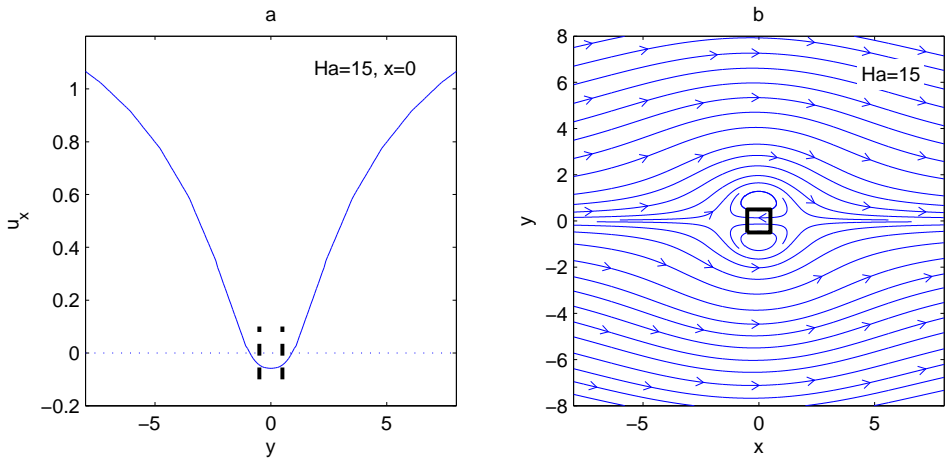}
\end{center}
    \caption{\label{Fig:Ha015_2D}
    Streamwise velocity $u_x$ along spanwise direction at $x=0$ ($a$) and
    flow streamlines ($b$) at $H=15$ for 2D MHD flow.
     On ($b$) one distinguishes two magnetic vortices.
    Bold lines denote borders of the magnet.}
\end{figure*}

\begin{figure*}[]
\begin{center}
    \includegraphics[width=12.5cm, angle=0, clip=on]{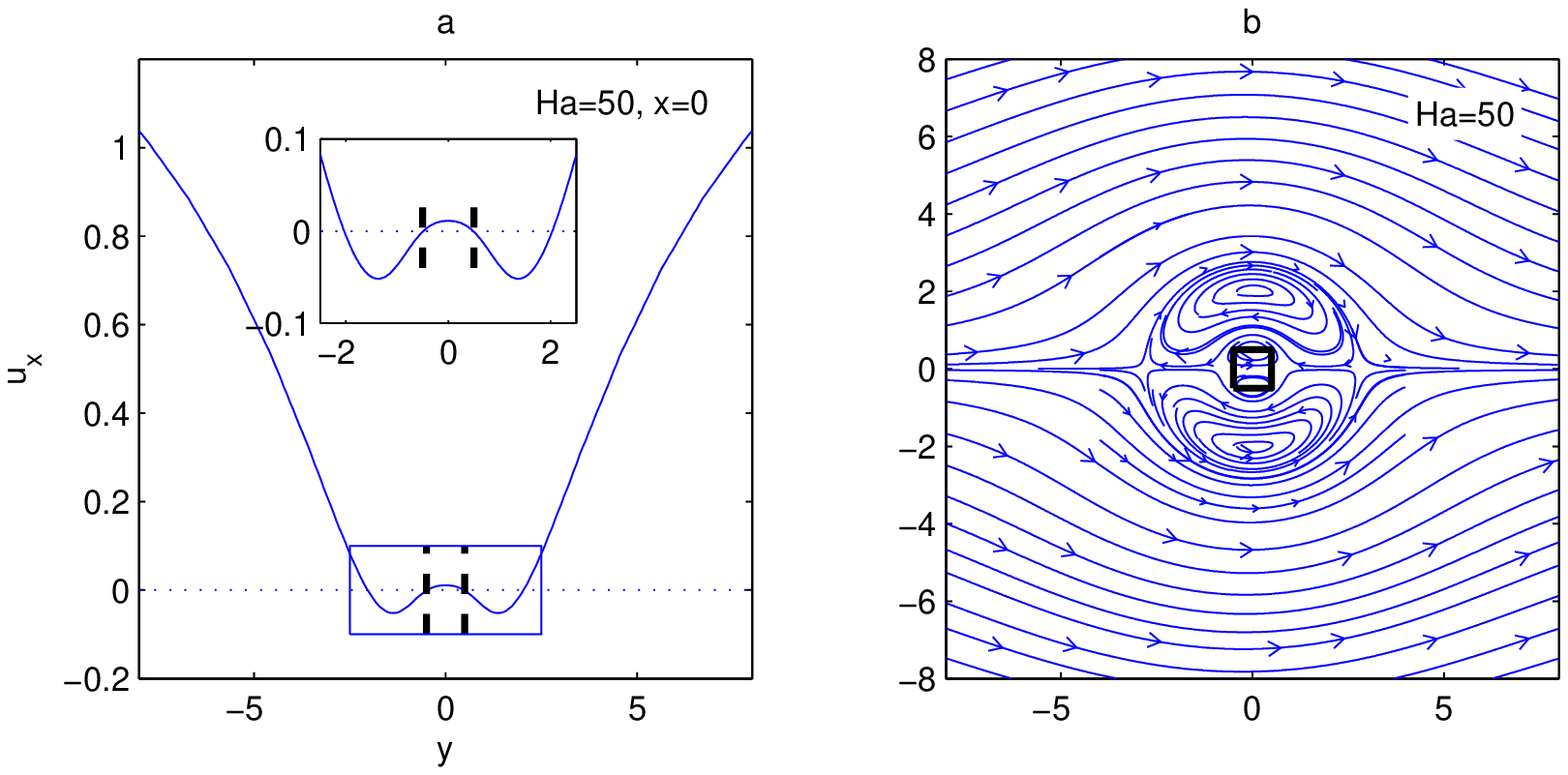}
\end{center}
    \caption{\label{Fig:Ha050_2D}
    Streamwise velocity $u_x$ along spanwise direction at $x=0$ ($a$) and
    flow streamlines ($b$) at $H=50$ for 2D MHD flow.
    Insertion on ($a$) is the zoom for $-1.5\le y\le 1.5$. On
    ($b$) one distinguishes four vortices aligned along the $y$
    direction at $x=0$.
    Bold lines denote borders of the magnet.}
\end{figure*}

\begin{figure*}[]
\begin{center}
    \includegraphics[width=12.5cm, angle=0, clip=on]{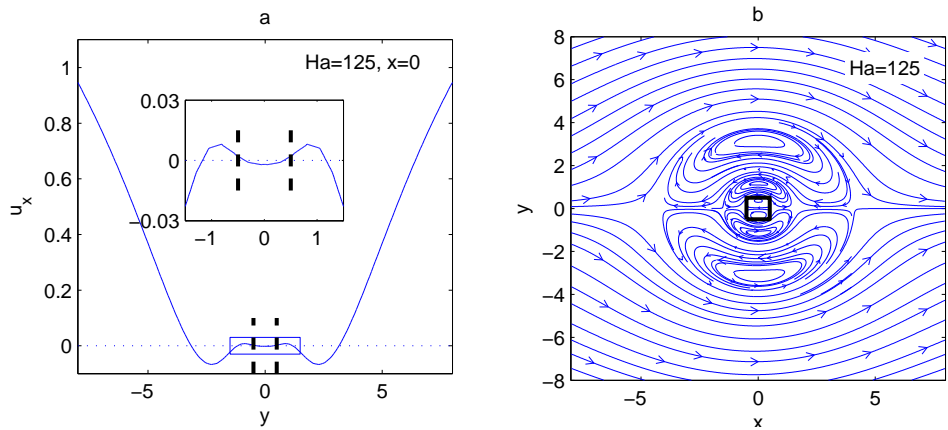}
\end{center}
    \caption{\label{Fig:Ha125_2D}
    Streamwise velocity $u_x$ along spanwise direction at $x=0$ ($a$) and
    flow streamlines ($b$) at $H=125$ for 2D MHD flow.
    Insertion on ($a$) is the zoom for $-1.5\le y\le 1.5$. On
    ($b$) one distinguishes six vortices aligned along the $y$
    direction at $x=0$.
    Bold lines denote borders of the magnet.}
\end{figure*}

To consider in detail the different flow regimes shown
cumulatively in Fig.~\ref{Fig:u_summ_2DRe005}, we have labelled
the turning points, where $u_{center}$ changes its sign, with the
letters A,B,C. Figures~\ref{Fig:Ha015_2D}-\ref{Fig:Ha125_2D}
illustrate below three nontrivial flow regimes through two plots:
the streamwise velocity profile along the central spanwise line,
$x=0$, and flow streamlines.

The first regime, corresponding to the  $Ha$ range from zero up to
the first turning point, A, is characterized by a weak breaking
Lorentz force. There is no recirculation, so we have considered
this regime as trivial, and did not illustrate it further by any
figure. It is marked as regime I and represented by Fig.~7 in the
paper by Cuevas \textit{et al.}
\cite{Cuevas:Smolentsev:Abdou:PRE:2006}.

The second  regime is for the $Ha$ interval between the first
turning point, A, and the second, B. It is characterized by a
Lorentz force that is already strong enough, to reverse the flow
between the magnetic poles. This results in two magnetic vortices
as illustrated by Fig.~\ref{Fig:Ha015_2D} of the current paper and
Fig.~8 of \cite{Cuevas:Smolentsev:Abdou:PRE:2006}. These vortices
are analogous to those in 3D results, with the difference that in
the 3D simulation the flow particles captured by the magnetic
vortices move helically towards the top and bottom walls to
dissipate upstream kinetic energy \cite{Votyakov:JFM:2007}, while
in the 2D system they always follow rotation along closed contour
lines of the stream function.

In the third regime, for the $Ha$ range between turning points B
and C, the reversal of the upstream flow occurs in the shear layer
rather than in the center of the magnetic obstacle. This can be
explained by the fact that,  at these $Ha$, the flow in the core
is hindered, and the shear layer between the core and the
bypassing flows is large to accommodate magnetic vortices. When
this happens, the requirement of flow continuity, $\nabla
\mathbf{u}_\perp=0$, turns the velocity in the center of the
obstacle to positive values. As a result, the four vortex, aligned
along the central spanwise line, appear as shown in
Fig.~\ref{Fig:Ha050_2D} (see also Fig.~9 of
\cite{Cuevas:Smolentsev:Abdou:PRE:2006}).

When $Ha$ increases  further, after the turning point C, the shear
layer is extended further as well. So the whole elongated vortex
structure increases in length, and two vortices, that were closest
to the center of the stagnant core, are pushed out from the core.
Again, to satisfy flow continuity this initiates the weaker
counter-rotating vortices in the core. In this regime, there are
altogether six aligned vortices shown in Fig.~\ref{Fig:Ha125_2D}.
This recirculation pattern is new and has not been reported
earlier.

Now, we can offer the following qualitative description of the
creeping 2D MHD flow around a magnetic obstacle. As $Ha$ increases
more and more, additional vortices appear in the expanded shear
layer. These vortices have decaying intensity towards to the
magnetic gap and are alternating-signed, starting from the very
first vortex, counted from the bypassing flow, where the flow is
reversed. This results in vanishing velocity oscillations in the
center of the magnetic gap, Fig.~\ref{Fig:u_summ_2DRe005}.

The question arises whether the aligned multi-vortices appear in a
creeping 3D MHD flow. In the latter case, the largest $N$ under
consideration was 1000 and two magnetic vortices were observed
only. The four-vortex aligned structure that was clearly presented
in our 2D results, occurred at $Ha \approx 50$, which corresponds
to ($Re=0.05$) $N=Ha^2/Re=50000$, much larger than the maximum
$N=1000$ used in the 3D simulations. On the other hand, the
velocities are of magnitude $10^{-3}-10^{-2}$ at the center of the
magnetic gap, so one would have to employ a very accurate 3D
numerical code to verify the effect discussed.

Physically, the crucial difference  between 2D and 3D vortices is
the secondary flow in the transverse direction. This dissipates
kinetic energy of the upstream flow inside viscous boundary layers
located on no-slip top/bottom walls of the duct. On the other
hand, the secondary flow variation, $\partial_{z}u_z$, is a mass
source/sink adjusted to keep flow continuity, $\nabla_\perp
\mathbf{u}_\perp=-\partial_{z}u_z$. Obviously in 2D cases, the
dissipation by the secondary flow is absent and the flow is
obliged to satisfy 2D continuity, $\nabla_\perp
\mathbf{u}_\perp=0$. Then, the possible way to dissipate energy
and the flow be continuous in 2D is to create a weaker
counter-rotating vortex nearby a strong vortex, the phenomena
observed in Fig.~\ref{Fig:Ha050_2D} and \ref{Fig:Ha125_2D}.

\section{Core of the magnetic obstacle and controlled
production of vorticity.}

It is worth to discuss how a magnetic obstacle can be used to
control vorticity and turbulence. On the one hand, the magnetic
field damps turbulent pulsations in the core of the obstacle by
freezing any kind of mass and electric transfer. On the other
hand, the shear layer between the core of the obstacle and the
rest of the flow generates a transverse vorticity. The results
presented herein were obtained at low $Re$ numbers, and all the
vortices were confined in the shear layer. However, it is possible
to imagine the following two scenarios. The first situation is to
destroy the initial core by increasing the $Re$ number, that is by
turbulating the flow, at fixed magnetic field parameters; and the
second situation is to increase the magnetic field strength so as
to manifest the core, that is to laminarize the core, while the
$Re$ number remans high and the rest of the flow is turbulent.

In the  trivial turbulating scenario, the interaction parameter
$N=Ha^2/Re$ is initially high because of the $Re$ number being low
while the $Ha$ number is moderate\footnote{A moderate $Ha$ number
range was also used in this paper in order to avoid numerical
difficulties related to the resolution of Hartmann layers.} Then,
one increases the $Re$, which could be equivalent to increasing
the inlet flow rate. Obviously, since $Ha$ is fixed, a higher $Re$
implies a lower $N$, hence, the core of the magnetic obstacle
becomes more penetrable, so the lateral magnetic vortices will
shift toward  each other due to the unfreezing of the core. At
some threshold $Re$ number, when the inertia of the flow becomes
sufficiently high to produce a stagnant region behind the core of
the obstacle, one will have to observe the six-vortex pattern
shown schematically in Fig.~\ref{Fig:IntroductoryMO}. Then, an
even higher $Re$ number will result in the detachment of the
attached vortices (the third pair of vortices in
Fig.~\ref{Fig:IntroductoryMO}) which is analogous to the classical
process of the detachment of the attached vortices past solid
cylinder \cite{Votyakov:PoF:2009}. Finally, at the highest $Re$,
the interaction parameter $N$ will become so small that the
internal recirculation structure of the magnetic obstacle will
disappear, and the flow will become turbulent as in the ordinary
hydrodynamics.

In the laminarizing scenario, by taking initially a high $Re$
number and keeping it fixed, one increases the $Ha$ number by
imposing, for instance,  an external local magnetic field with
higher and higher intensity. This increases the interaction
parameter $N$, so the core of the obstacle must manifest itself by
first showing  the same six-vortex recirculation as
aforementioned. In practice however, it might be difficult for all
six vortices to be stable at very high $Re$, because in this case
the rest of the flow is turbulent, so the wake of the obstacle
oscillates and can potentially destroy the stability of the
connecting and attached vortices (the second and the third pair of
the vortices in Fig.~\ref{Fig:IntroductoryMO}). Nevertheless, the
lateral magnetic vortices (the first pair of the vortices in
Fig.~\ref{Fig:IntroductoryMO}) must manifest themselves clearly at
any $Re$ provided that $N$ is large. These vortices will be
distinctly seen at the beginning of the turbulent wake. A further
increase of the  $Ha$  does laminarize and freeze the core of the
obstacle. The magnetic vortices move away from each other and
adjust their position in the lateral shear layers. Owing to the
high $Re$ number, the recirculation confined in the shear layer
possesses an excess amount of kinetic energy that is dissipated
downstream in the wake of the obstacle.

In the case of a solid obstacle one has little control over the
shear layer at high $Re$ because there is one parameter only, i.e.
$Re$. Also, the solid obstacle is impenetrable and could not fit
itself to the flow. In the case of the magnetic obstacle one has
an opportunity to govern the detachment process because there are
two additional parameters at a given  $Re$: the $Ha$ number, for
instance, the magnetic field intensity controlling the degree of
permeability of the magnetic obstacle; and the degree of lateral
heterogeneity of the external magnetic field controlling the width
of the lateral shear layer. Thus, a controlled intensity and shape
of a magnetic obstacle is a model lab to produce and study
vorticity generation and turbulent phenomena.

\section{Conclusions.}

3D numerical simulations are reported for a liquid metal flow
subject to a strong external heterogenous magnetic field. The
simulations shed light on the process of formation of the core of
the magnetic obstacle when the interaction parameter $N$ is large.
The core is surrounded by deformed magnetic vortices located in
the shear layer. Inside the core, there is no mass and electric
transfer, i.e. at high $N$, the magnetic obstacle is analogous to
a solid hydrodynamical obstacle.

The series of 2D simulations for a creeping MHD flow demonstrated
that there still exists a weak recirculation in the stagnant core
in 2D cases even at very high $Ha$. The flow regime can be
represented as an elongated recirculation composed of many (even
number) sign-alternating vortices aligned along the line crossing
the center of the magnetic gap in the spanwise direction. The
intensity of vortices vanishes by approaching the core of the
magnetic obstacle, therefore, the  principal conclusion derived
from 3D results remains valid: there develops a frozen core of the
magnetic obstacle when  $Ha$ grows toward infinity. By studying
the characteristics of magnetic obstacles in laminar flow, we have
been able to provide conjectures on the formation and destruction
of magnetic obstacle in turbulent flow.

\section{ACKNOWLEDGEMENTS}
This work has been performed under the UCY-CompSci project, a
Marie Curie Transfer of Knowledge (TOK-DEV) grant (Contract No.
MTKD-CT-2004-014199). This work was also partially funded under a
Center of Excellence grant from the Norwegian Research Council to
the Center of Biomedical Computing.


\bibliography{./../../Bibtex/mhd}

\begin{thebibliography}{27}
\providecommand{\natexlab}[1]{#1}

\bibitem[1]{Votyakov:PRL:2007}
E.V. Votyakov, Y. Kolesnikov, O. Andreev, E. Zienicke, and A. Thess, {\itshape
  Structure of the wake of a magnetic obstacle}, Phys. Rev. Lett. 98 (2007), p.
  144504.

\bibitem[2]{Davidson:Review:1999}
P. Davidson, {\itshape Magnetohydrodynamics in {M}aterials {P}rocessing},
  Annual Review of Fluid Mechanics 31 (1999), pp. 273--300.

\bibitem[3]{Thess:Votyakov:Kolesnikov:2006}
A. Thess, E.V. Votyakov, and Y. Kolesnikov, {\itshape {L}orentz {F}orce
  {V}elocimetry}, Phys. Rev. Lett. 96 (2006), p. 164501.

\bibitem[4]{Shercliff:book:1962}
J.A. Shercliff {\itshape The theory of electromagnetic flow-measurement},
  Cambridge University Press, 1962.

\bibitem[5]{Roberts:1967}
P.H. Roberts {\itshape An introduction to {M}agnetohydrodynamics},    Longmans,
  Green, New {Y}ork, 1967.

\bibitem[6]{Moreau:book:1990}
R. Moreau {\itshape Magnetohydrodynamics},    {K}luwer {A}cademic {P}ublishers,
  Dordrecht, 1990.

\bibitem[7]{Davidson:book:2001}
P.A. Davidson {\itshape An introduction to {M}agnetohydrodynamics},
  Cambridge University Press, 2001.

\bibitem[8]{Moffatt:1967}
H.K. Moffatt, {\itshape On the suppression of turbulence by a uniform magnetic
  field}, J. Fluid. Mech. 28 (1967), pp. 571--592.

\bibitem[9]{Sommeria:Moreau:MHDturblulence:1982}
J. Sommeria, and R. Moreau, {\itshape Why, how, and when, {MHD} turbulence
  becomes two-dimensional}, J. Fluid. Mech. 118 (1982), pp. 507--518.

\bibitem[10]{Davidson:1997}
P.A. Davidson, {\itshape The role of angular momentum in the magnetic damping
  of turbulence}, J. Fluid. Mech. 336 (1997), pp. 123--150.

\bibitem[11]{Votyakov:PoF:2009}
E.V. Votyakov, and S.C. Kassinos, {\itshape On the analogy between streamlined
  magnetic and solid obstacles}, Phys. Fluids 21 (2009), pp. 097102--11.

\bibitem[12]{Gelfgat:Peterson:Sherbinin:1978}
Y.M. Gelfgat, D.E. Peterson, and E.V. Shcherbinin, {\itshape Velocity structure
  of flows in nonuniform constant magnetic fields 1. Numerical calculations.},
  Magnetohydrodynamics 14 (1978), pp. 55--61.

\bibitem[13]{Gelfgat:Olshanskii:1978}
Y.M. Gelfgat, and S.V. Olshanskii, {\itshape Velocity structure of flows in
  non-uniform constant magnetic fields. II. Experimental results.},
  Magnetohydrodynamics 14 (1978), pp. 151--154.

\bibitem[14]{Cuevas:Smolentsev:Abdou:Pamir:2005}
S. Cuevas, S. Smolentsev, and M. Abdou, {\itshape Vorticity generation in
  non-uniform mhd flows}, in {\itshape Proceedings of the Joint 15th Riga and
  6th {PAMIR} International Conference. Fundamental and Applied {MHD}}, Vol.
  ~1, Riga, Jurmala, Latvia, June 27-July 1, 2005, pp. 25--32.

\bibitem[15]{Cuevas:Smolentsev:Abdou:2006}
S. Cuevas, S. Smolentsev, and M. Abdou, {\itshape On the flow past a magnetic
  obstacle}, J. Fluid. Mech. 553 (2006), pp. 227 -- 252.

\bibitem[16]{Cuevas:Smolentsev:Abdou:PRE:2006}
S. Cuevas, S. Smolentsev, and M. Abdou, {\itshape Vorticity generation in
  creeping flow past a magnetic obstacle}, Phys. Rev. E 74 (2006), p. 056301.

\bibitem[17]{Votyakov:JFM:2007}
E.V. Votyakov, E. Zienicke, and Y. Kolesnikov, {\itshape Constrained flow
  around a magnetic obstacle}, J. Fluid. Mech. 610 (2008), pp. 131--156.

\bibitem[18]{Molokov:Reed:Fusion:2003}
S. {Molokov}, and C.B. {Reed}, {\itshape {Parametric Study of the Liquid Metal
  Flow in a Straight Insulated Circular Duct in a Strong Nonuniform Magnetic
  Field}}, Fusion Science And Technology 43 (2003), pp. 200--216.

\bibitem[19]{Alboussiere:2004}
T. Alboussiere, {\itshape A geostrophic-like model for large {H}artmann number
  flows}, J. Fluid. Mech. 521 (2004), pp. 125--154.

\bibitem[20]{Kumamaru:etal:2004}
H. Kumamaru, S. Kodama, H. Hirano, and K. Itoh, {\itshape Three-Dimensional
  Numerical Calculations on Liquid-Metal Magnetohydrodynamic Flow in
  Magnetic-Field Inlet-Region}, Journal of Nuclear Science and Technology 41
  (2004), pp. 624--631.

\bibitem[21]{Kumamaru:etal:2007}
H. Kumamaru, K. Shimoda, and K. Itoh, {\itshape Three-Dimensional Numerical
  Calculations on Liquid-Metal Magneto-hydrodynamic Flow through Circular Pipe
  in Magnetic-Field Inlet-Region}, Journal of Nuclear Science and Technology 44
  (2007), pp. 714--722.

\bibitem[22]{ni_current_2007}
M. Ni, R. Munipalli, P. Huang, N.B. Morley, and M.A. Abdou, {\itshape A current
  density conservative scheme for incompressible {MHD} flows at a low magnetic
  Reynolds number. Part {II:} On an arbitrary collocated mesh}, Journal of
  Computational Physics 227 (2007), pp. 205--228.

\bibitem[23]{Votyakov:TSFP6:2009}
E.V. Votyakov, and S.C. Kassinos, {\itshape Core of the magnetic obstacle}, in
  {\itshape The sixth International Symposium on Turbulence and Shear Flow
  Phenomena, Vol.II}, Seoul, Repeublic of Korea, 2009, pp. 703--707.

\bibitem[24]{Votyakov:Zienicke:FDMP:2006}
E.V. Votyakov, and E. Zienicke, {\itshape Numerical study of liquid metal flow
  in a rectangular duct under the influence of a heterogenous magnetic field},
  Fluid Dynamics \& Materials Processing 3 (2007), pp. 97--113.

\bibitem[25]{Griebel:book:1995}
M. Griebel, T. Dornseifer, and T. Neunhoeffer {\itshape Numerische
  Str{\"o}mungssimulation in der Str{\"o}mungsmechanik},    Vieweg Verlag,
  Braunschweig, 1995.

\bibitem[26]{Kulikovskii:1968}
A. Kulikovskii, {\itshape Slow steady flows of a conducting fluid at high
  Hartmann numbers}, Izv. Akad. Nauk. SSSR Mekh. Zhidk. i Gaza  (1968), pp.
  3--10.

\bibitem[27]{Debnath:1973}
L. {Debnath}, {\itshape On {E}kman and {H}artmann Boundary Layers in a Rotating
  Fluid}, Acta Mechanica 18 (1973), pp. 333--341.

\end{thebibliography}
\bibliographystyle{tJOT}

\end{document}